\documentclass[twocolumn,pra,showpacs]{revtex4}

\usepackage{bm}
\usepackage{amsmath}
\usepackage{graphicx}

\begin{document}

\title{All-band Bragg solitons and $cw$ eigenmodes}

\author{A.~E.~Kaplan}
\email{alexander.kaplan@jhu.edu}
\affiliation{Dept.\ of Electrical and Computer Engineering,
Johns Hopkins University, Baltimore, MD 21218}

\date{\today}

\begin{abstract}
We found an amazingly simple general "all-band" intensity profile of
bandgap (Bragg) solitons for arbitrary parameters of 
spatially-periodic nonlinear systems,
similar to  those of multi-frequency stimulated Raman scattering,
in particular the so called Lorentzian-profile solitons.
We also found nonlinear eigen-modes of such system
that propagate without energy exchange between waves.
\end{abstract}


\preprint{To appear in \emph{Phys.~Rev.~A}}

\maketitle
%-------------------------------------------------------------------------
\section{Introduction}
\label{sec1}

In nonlinear optics, solitons, 
both temporal and spatial, 
became familiar objects, which
can be originated by quite a few 
fundamental nonlinear processes.
The ones of interest here are
so called gap (or Bragg) solitons [1-5]
(for review, see e. g. [6])
that are due to the interaction 
of light with spatially-periodic 
nonlinear structures, whereby
counterpropagating waves are
strongly coupled to each other $via$
distributed back-and-forth Bragg reflection.
The Bloch theory of $linear$ 
periodic structures predicts
the existence of "prohibited"
zones -- bandgaps -- where
due to resonances between incident wavelength
and the period of spatial modulation,
the incident wave cannot propagate
for long, and the structure 
becomes almost fully reflective;
those are essentially 
multilayered Bragg reflectors.
The spectral width of such
a bandgap is proportional to the contrast
between refractive indices of 
constituent materials (or waveguide).
When these refractive indexes also depend on
the light intensity (e. g. due to Kerr-like nonlinearity -- 
self-focusing if positive, self-defocusing if negative),
the system may exhibit a rich host
of nonlinear effects,
some of the most interesting being
gap solitons that emerge as a critical
phenomenon, whereby the light with its frequency
being inside the bandgap, "pushes"
the bandgap edge away and carves 
conditions for itself to penetrate 
deep into the structure by forming gap solitons.

A new property of bandgap solitons,
predicted first in [1], which
gave them a distinct place amongst the soliton crowd,
was that they may exist as standing 
(immobile, stationary) field objects.
In fact, they might be viewed 
as a simplest, 1D-self-trapping of light.
It was demonstrated then [2] that under certain
conditions those solitons could be 
similar to familiar nonlinear solitons
in regular fibers [3] and 2D-self-focusing [4]
and be governed by a rescaled
cubic Schr\"{o}dinger equation.

In further development, it has been shown [5] that 
Bragg systems can support even more general gap solitons
that have non-zero but very slow group velocity,
giving rise to nonlinear "slow light" objects.
The slow bandgap (SBG) solitons have later
been observed experimentally in [7], 
and immobile solitons -- in [8].
Gap solitons have by now been shown
to emerge in many other bandgap systems,
such as e. g. in a Bose-Einstein gas [9]
and near critical point in underdense plasma
due to relativistic nonlinearity of electrons [10].
While SBG-solitons have been shown [5] 
to exist within a linear bandgap, 
an approximate analytical solution in [5]
have been found  only at a close vicinity of the bandgap edge
(which in the case of self-focusing nonlinearity
we call here a "blue edge", see Section IV below).
Its intensity profile was again
similar to nonlinear Schr\"{o}dinger equation soliton.
(As one can see below, Section IV, it is
the weakest and longest soliton.)
A greatly important development in the field was a proof [11] 
(based on  generalization of a so called "massive Thirring model",
see references in [11])
that related coupled nonlinear partial differential equations 
are fully integrable, which in particular allowed for a 
complex solution for stationary SBG-solitons
with an arbitrary detuning within a bandgap.
%\vspace{-.7in} 

In this paper we show that 
the family of intensity profiles of all
the solitons within the entire
Bragg bandgap are described by
a simple elegant general formula (see Eqs. (3.6) and (4.1) below),
which may be called an all-band Bragg soliton,
derived by us based on
a standard Bragg wave equation (Section II)
and directly looking for a solution
that satisfies the conditions of a fixed group 
velocity and bright-soliton asymptotic (Section III).
It can be shown that
these solutions are consistent with the general results [11],
yet their analytical simplicity and transparency 
related to the use of intensities and phases as output variables,
allowed for easy visualization
and analysis of the soliton intensity profiles, in particular in
the less studied, and perhaps the most interesting
area near red-edge of the bandgap
(for a positive Kerr-nonlinearity), where
they assume an extremely simple yet unusual
for solitons Lorentzian profile, see (4.5) below,
which was missing in [11];
they are similar to a limiting case
of Raman scattering solitons [12].
This approach also brings up all the relevant invariants of motion.
We analyzed in detail
both moving (Section IV) and stationary solitons (Section V).
Using a similar approach for $cw$, $non-soliton$,
solution, we found $cw-eigenmodes$ of
the system, in particular in a ring Bragg reflector (Section VI),
and discussed the ramification
and possible applications of our results (Section VII).

\section{Wave propagation model: Bragg reflection + Kerr-like nonlinearity}
\label{sec2}

Considering a 1D-model (e.g. as in an optical
fiber), we assume that the periodic grating in it is formed 
by the modulation of the refractive index, $n$
(this can also be done by slight periodic tapering of fiber,
which can be dealt with by a similar math-description.)
In a uniform, unmodulated ($n=const$) fiber the electric field $E$ 
propagation is governed by a regular wave equation
\begin{equation}
{\partial^2 E } / { \partial z^2 } -  
( n^2 / c^2 )  {\partial^2 E } / { \partial t^2 } = 0
\tag{1.1}
\end{equation}
which in low-dispersion case can be decomposed into the set
of two first-order equations for 
"forward", $E_1$, and "backward", $E_2$, traveling waves as:
\begin{equation}
( -  1 )^j {\partial E_j } / { \partial z } - 
( n / c )  {\partial  E_j } / { \partial t } = 0 ; \ \ \      j = 1 , 2
\tag{1.2}
\end{equation}
whose $\omega$-monochromatic radiation solution is 
$E_j = A_j$ $ \exp [ - i ( - 1 )^j k z - i \omega t ] / 2  +  c.c.$, 
with $A_j = const$, and
$k = \omega / c n  = 2 \pi / n \lambda$,
$n$ is an unperturbed refractive index, 
and $\lambda$ is a free space wavelength.
(As is common in the bandgap theory,
we will neglect here the intrinsic
dispersion of $n ( \omega )$ in a fiber, 
since the bandgap dispersion
due to index modulation greatly exceeds that
of a regular waveguide [5].)

A spatially modulated fiber making
a periodic grating with a period $L_B$, 
has its resonant Bragg wavenumber, free space wavelength, 
and frequency respectively as
$k_B = \pi / L_B$,
$\lambda_B =  2 n_0 L_B$ and  
$\omega_B = k_B c / n_0 $;
a normalized half-width of a linear bandgap around $\omega_B$
is $\mu = \Delta n / n_0 \ll  1$, where
$\Delta n$ is the index modulation amplitude.
No radiation with $\Delta^2  \le \mu^2 $
is allowed  then to propagate in  a sufficiently long system;
here $\Delta  = \omega / \omega_B -  1$ is a normalized Bragg detuning.
Furthermore, in the presence of 
a Kerr-like nonlinearity, refractive index
depends on the intensity of 
light as $n = n_0 + n_K  E^2$,
where $n_K$ is a coefficient due to $\chi^{(3)}$ 
(Kerr) nonlinearity.
The total refractive index can be written then as
\begin{equation}
n = n_0 + \Delta n \cos ( 2 k_B z )  + n_K  E^2
\tag{1.3}
\end{equation}
Generalizing (1.1) to have the index $n$
as in (1.3), and seeking its solution as
the sum of traveling waves, $E = E_1 + E_2$, with
\begin{equation}
E_j =  A_j ( t , z ) \exp [ -i \omega_B 
( 1 + \Delta ) t  -  
%\notag
%\end{equation}
%\begin{equation}
i ( - 1 )^j k_B z ] / 2  + c.c. ;
%\ \ \ \    j = 1, 2
\tag{1.4}
\end{equation}
with $j = 1, 2$,
presuming the envelopes $A_j$ to vary slowly 
in time and space, as
$ | \Delta n |,  | n_K  E^2 | \ll n_0$,
and neglecting higher-harmonics generation
(a common approach [1-6] in nonlinear Bragg reflection
because of large difference in phase velocities), 
we obtain truncated equations of evolution 
being nonlinear+modulated counterparts of
(1.2) for those envelopes as:
\begin{equation}
%i  \left[ - ( -  1 )^j {{\partial A_j}  \over   {k_B \partial z }} + 
%{{ \partial A_j}  \over  {\omega_B  \partial t }} \right] + 
i  \left[ - ( -  1 )^j {\partial A_j}  / {k_B \partial z } + 
{{ \partial A_j}  /  {\omega_B  \partial t }} \right] + 
A_j \Delta  +  \mu A_{{3} - j}  + 
\notag
\end{equation}
\begin{equation}
{ ( {n_K} / {n_0} ) } (  | A_j |^2 + 
2  | A_{{3} - j} |^2 )   A_j = 0 ;
\ \ \ \    j = 1, 2
\tag{1.5}
\end{equation}
where the factor $2$ in the nonlinear term  is originated
by intensity-induced non-reciprocity [13].
In linear case Eq. (1.5) is similar to
coupled equations for counterpropagating modes 
in a Bragg reflector [14].

To normalize (1.5), we use the scales for intensity,
$ E_{NL}^2 =  \Delta n / n_K $, 
distance, $z_{NL} = 1 / \mu k_B$,
and time, $t_{NL} =  1 / \mu \omega_B$.
Introducing then dimensionless envelopes,
$a_j = A_j / E_{NL}$, time $\tau  =  t / t_{NL} $, 
and distance $\zeta = z / z_{NL}$, 
we rewrite (1.5) as:
\begin{equation}
i  \left[ - ( -  1 )^j { \partial a_j  \over   \partial \zeta } + 
{ \partial a_j  \over  \partial \tau } \right] + 
a_j \delta  +  a_{{3} - j}  +  
\notag
\end{equation}
\begin{equation}
s_K \left( u_j^2 + 
2 u_{{3} - j}^2 \right)   a_j = 0 ;    
\ \ \ \
u_j^2 \equiv   | a_j |^2
\tag{1.6}
\end{equation}
where $s_K = sign ( n_K / \Delta n )$.
For a fixed $s_K$, 
only one controlling parameter,
a normalized detuning 
$\delta = \Delta / \mu =$ 
$( \omega / \omega_B -  1 ) n_0 / \Delta n$,
is left in (1.6).
A linear band is constituted then by the condition
$\delta^2  \le 1$.

\section{Waves with fixed group velocity}
\label{sec3}
As a next step let us find the solution of (1.6) 
for the class of modulated coupled waves 
with a fixed group velocity, 
$\beta = {\bf v}_{gr} n_0 / c$, ($\beta^2 \le 1$);
the solitons will be part of that family.
To that end, we assume that in the frame moving with that velocity,
i. e. "comoving frame", the fields are time-independent.
Introducing then time-space variables in that frame as
$\zeta_{\beta} = \zeta - \beta \tau$, $\tau_{\beta}  = \tau$,
we are looking for the solutions
with  $ \partial / \partial \tau_{\beta} = 0$
to derive ordinary nonlinear differential equations
\begin{equation}
-  i  [ \beta +  ( - 1 )^j ] a_j ^\prime   + 
\delta  a_j +  a_{{3} - j} +  s_K 
(  u_j^2  +  2 u_{{3} - j}^2 )   a_j = 0
\tag{2.1}
\end{equation}
where "prime" stands for $d / d \zeta_{\beta}$.
To elucidate direct results for the intensity profiles,
from this point, as different from [11],
we will be using an
"amplitude\&phase" approach similar 
to our previous work [15,10], 
which greatly simplifies the calculations.
To that end, we write $a_j = u_j \exp ( i \phi_j )$ 
with $u_j$ and $\phi_j$ -- real,
rewrite (2.1) as
\begin{equation}
- [ \beta + ( -  1 )^j ]  ( i {u_j} ^\prime  - 
{\phi_j} ^\prime u_j ) +  
u_j [ \delta +  s_K ( u_j^2  +  2 u_{{3} - j}^2 ) ] + 
\notag
\end{equation}
\begin{equation}
u_{3 - j} e^{i ( \phi_{3 - j} - \phi_j ) } = 0
\tag{2.2}
\end{equation}
and separate real and imaginary parts in Eq. (2.2):
\begin{equation}
[ 1 +  ( -  1 )^j  \beta ] {u_j} ^\prime  =  
u_{{3} - j} \sin ( \phi_1 -  \phi_2 ) ;
\tag{2.3}
\end{equation}
\begin{equation}
u_j \ [ ( -  1 )^j + \beta ] {\phi_j} ^\prime  - 
[ \delta +  s_K ( u_j^2  +  2 u_{{3} - j}^2 ) ] \ = 
\notag
\end{equation}
\begin{equation}
- u_{{3} - j}  \cos ( \phi_1 -  \phi_2 ) .
\tag{2.4}
\end{equation}
Using eq (2.3) for both $j = 1$ and $j = 2$,
we have $( u_1^2  -  u_2^2 ) ^\prime  = 
\beta ( u_1^2  +  u_2^2 ) ^\prime$, 
hence the invariant of motion for this class of solutions:
\begin{equation}
( u_1^2 -  u_2^2 )  - 
\beta  ( u_1^2 +  u_2^2 )  = D  = inv
\tag{2.5}
\end{equation}
Using this in (2.3) and (2.4) and
introducing "combined" variables
\begin{equation}
S =  u_1^2 +  u_2^2 ; \ \ \
\Phi =  \phi_1  - \phi_2 ; \ \ \
%\notag
%\end{equation}
%\begin{equation}
\Sigma =  \phi_1  + \phi_2 ;     
\ \ \  P = u_1 u_2
\tag{2.6}
\end{equation}
we see that their spatial dynamics is governed by equations:
\begin{equation}
S ^\prime = {{4 P \sin \Phi}  \over  { 1 -  \beta^2 }} ;
\ \ \ \
P ^\prime = {S ( 1 -  \beta^2 ) -  {\beta D }  
\over  1 -  \beta^2 }  \sin \Phi
\tag{2.7}
\end{equation}
\begin{equation}
\Phi ^\prime = - {{ 2 ( \delta +  s_K S )}   + 
[ S ( 1 -  \beta^2 ) -  {\beta D } ]  
( s_K -  {\cos \Phi} / P )  
\over  1 -  \beta^2 }
\tag{2.8}
\end{equation}
\begin{equation}
\Sigma ^\prime = -  { 2 \beta ( \delta + s_K  S ) - 
D \left( s_K -  {\cos \Phi} / P \right)  
\over  1 -  \beta^2 }
\tag{2.9}
\end{equation}
From eqns (2.7) we obtain phase space
equation for $P , S$ independent of  $\Phi$ and $\Sigma$,
which greatly simplifies further calculations
by separating those variables:
\begin{equation}
{ d S   \over d P } =  { 4 P   \over S ( 1 -  \beta^2 ) - \beta D }
\tag{2.10}
\end{equation}
whose solution, e. g. for $P ( S )$, is immediately found as
\begin{equation}
P ( S ) =   \sqrt { ( 1 -  \beta^2 )  ( S  - S_{min} )  
[ ( S  + S_{min} ) / 2 -   \beta D ] / 2 }
\tag{2.11}
\end{equation}
where $S_{min}$ is an integration constant chosen in such a way
that $P = 0$ when $S  = S_{min}$.
Having this result and using (2.8) and first eqn (2.7),
we obtain now a phase space eqns for $\Phi , S$, as
\begin{equation}
{ d \Phi   / d S } = 
\tag{2.12}
\end{equation}
\begin{equation}
- {{ 2 ( \delta +  s_K S )}  + 
[ S ( 1 -  \beta^2 ) -  {\beta D } ]  
[ s_K -  {\cos \Phi} / P ( S ) ]  
\over  4 P ( S ) \sin \Phi } 
\notag
\end{equation}

\section{All-band moving solitons}
\label{sec4}

The solution of (2.12) in the form e. g. $\Phi ( S )$,
being substituted in (2.7) and (2.8), 
after integration yields $S$, $P$, and $\Phi$ as
function of $\zeta_{\beta}$.
In general, all of them can be shown
to be periodic functions of the distance $\zeta_{\beta}$,
except for solitons.
The so called bright solitons
must then satisfy the condition,
\begin{equation}
u_j \to 0 \ \ \     at \ \ \ \      | \zeta |  \to  \infty
\tag{3.1}
\end{equation}
From (2.5) then $D = 0$, hence
$u_1^2 -  u_2^2  = \beta  ( u_1^2 +  u_2^2 )$.
From (2.10) and (2.11) we have also that $S_{min} = 0$ and
\begin{equation}
P = S \sqrt {1 -  \beta^2} / 2 ,
\tag{3.2}
\end{equation}
so that eqn (2.12) for variables $S$ and $\Phi$ reads now as:
\begin{equation}
{ d \Phi   \over d S} = - 
{2 \delta +  s_K S ( 3 - \beta^2 ) - 
2 \sqrt {1 -  \beta^2} \cos \Phi  \over  2 S \sin \Phi \sqrt {1 -  \beta^2}}
\tag{3.3}
\end{equation}
the solution of which is readily found as
$\cos \Phi = C / S +  [ \delta + S ( 3 -  \beta^2 ) / 4 ] /
\sqrt {1 -  \beta^2}$
where $C$ is yet another integration constant.
Due to (3.1) we set $C = 0$, so finally
\begin{equation}
\cos \Phi =  [ \delta +  s_K S ( 3 -  \beta^2 ) / 4 ] /
\sqrt {1 -  \beta^2}
\tag{3.4}
\end{equation}
Using (3.4) and (3.2) in the first eqn (2.7), we obtain a first
order eqn for one variable, $S ( \zeta_{\beta} )$, alone
\begin{equation}
S ^\prime =  2 S {{\sqrt { 1 -  \beta^2 - 
[ \delta +  s_K S ( 3 -  \beta^2 ) / 4 ]^2 }}  \over 
{ 1 -  \beta^2 }}
\tag{3.5}
\end{equation}
whose elegantly simple yet little familiar
in the soliton theory solution can be readily found:
\begin{equation}
S = {S_{pk}  ( 1 +  B )  \over   \cosh [ \alpha ( \zeta_{\beta} -  \zeta_0  ) ]   +  B } ; \ \ \    
B = {{ \delta s_K}   \over {\sqrt { 1 -  \beta^2} } }
\tag{3.6}
\end{equation}
where combined peak intensity, $S_{pk}$, and size-related 
parameter $\alpha$ are
\begin{equation}
S_{pk} =  { 4 \sqrt {1 - \beta^2 } ( 1  -  B )  \over 3  -  \beta^2 } ,   
\ \ \ \
\alpha =  {2 \sqrt { 1 - \beta^2 -  \delta^2}  \over  1 -  \beta^2} 
\tag{3.7}
\end{equation}
In (3.6) the soliton peak position, $\zeta_0$, is 
yet another integration constant; 
without loss of generality we will assume $\zeta_0 = 0$.

Amazingly, profiles (3.6) that suggest a broad control
of the shape and width of a soliton determined
by parameters $\alpha$ and $B$
(notice here that $B$ can be either negative or positive),
coincide with the soliton profiles
that show up in the multi-cascade
stimulated Raman scattering [12],
with completely different mechanism of their formation.
Apparently, such solitons are common
solutions for many situation and system
with interference between waves with
different wave-vectors: in those two examples either
co-propagating waves (as in Raman scattering),
or counter-propagating waves (as in Bragg scattering).

Individual waves intensities are now found $via$ (2.5) as
\begin{equation}
u_1^2 =  S ( 1 +  \beta ) / 2 ;  \ \ \ \ \ \ \ 
u_2^2 =  S ( 1 -  \beta ) / 2 .
\tag{3.8}
\end{equation}
Substituting (3.6) into (3.4), we have an expression for the phase 
difference $\Phi \equiv \phi_1 - \phi_2$:
\begin{equation}
\cos \Phi =  {  1 + B \cosh ( \alpha \zeta_{\beta} )  
\over B + \cosh ( \alpha \zeta_{\beta} )} ;
\ \ \ or
\notag
\end{equation}
\begin{equation}
\tan \Phi =  { \sqrt { 1 -  \beta^2  -  \delta^2}  
\sinh ( \alpha \zeta_{\beta} ) 
\over   \sqrt { 1 -  \beta^2}  [ 1 + B \cosh ( \alpha \zeta_{\beta} ) ] }
\tag{3.9}
\end{equation}
At the peak, $\zeta_{\beta} =  0$, we have
$\Phi = 0$, so the counterpropagating waves are 
of the same phase at that point, $\phi_1 = \phi_2$.
At $  \zeta_{\beta}  \to \infty$, we have
$ \tan \Phi  \to \sqrt { 1 -  \beta^2 - \delta^2} / \delta $.
At the exact resonance, $\delta  = 0$, 
$\Phi$ is reversed by $\pi$ as we go from
$\zeta_{\beta} = - \infty$ to $\zeta_{\beta} = \infty$.
Tedious but mundane calculations show that Eqns. (3.7) -- (3.9) 
are consistent with results [11] (see Eq. (6) in [11]);
in particular, the major soliton parameter $Q$
in [11] is related to the parameters 
used here $via$ $\cos Q = B$.
The simplicity and transparency of 
(3.7) -- (3.9) however, allows one to easily
analyze the soliton intensity profiles, in particular in
the less studied, and perhaps the most interesting 
area near red-edge of the bandgap
(for a positive Kerr-nonlinearity), where
they assume an extremely simple yet unusual 
for solitons Lorentzian profile, see (4.5) below.

Notice from (3.6) -- (3.8) that the 
bandgap for the moving solitons with a given
velocity, $\beta$, is narrower than
in the linear case: $\delta^2  < 1 - \beta^2$.
By the same token, near both the edges of a bandgap, 
$1  - \delta^2  \ll 1$,
only slow solitons are allowed, since then
$\beta^2 < 1 - \delta^2$.
On the other hand, in the middle of the bandgap, $\delta = 0$,
the solitons are allowed with the speed up 
to the maximum $\beta^2  = 1$.

\section{Stationary (immobile) solitons}
\label{sec5}

For immobile, or stationary, solitons,
$\beta  = 0$, we have $u_1^2 = u_2^2 = S / 2$
(hence the analogy to standing waves),
and the combined intensity profile, $S$, and phase $\Phi$ 
are found from (3.6) and (3.9) as
\begin{equation}
S = 
{ 4 ( 1 -  \delta^2 ) / 3  \over  \delta +  \cosh ( 2 \zeta \sqrt {1 - \delta^2 } )  } ;
\tag{4.1}
\end{equation}
\begin{equation}
\tan \Phi = {\sqrt {1 -  \delta^2} \sinh ( 2 \zeta \sqrt {1 - \delta^2 })  
\over   1 + \delta \cosh ( 2 \zeta \sqrt {1 - \delta^2 } ) }
\tag{4.2}
\end{equation}
Since this analytical solution in the entire 
bandgap, i. e. for any $\delta^2  \le 1$
is now available, one can analyze how it
evolves as the laser frequency is tuned
from upper (blue) edge of the bandgap, $\delta = 1$, to the lower
(red) edge, $\delta = - 1$.
For example, close to the blue edge, 
$0 < 1 - \delta \equiv \Delta \delta \ll 1$,
we have a low-intensity long pulse 
\begin{figure}
\includegraphics[angle=270,width=3.5in]{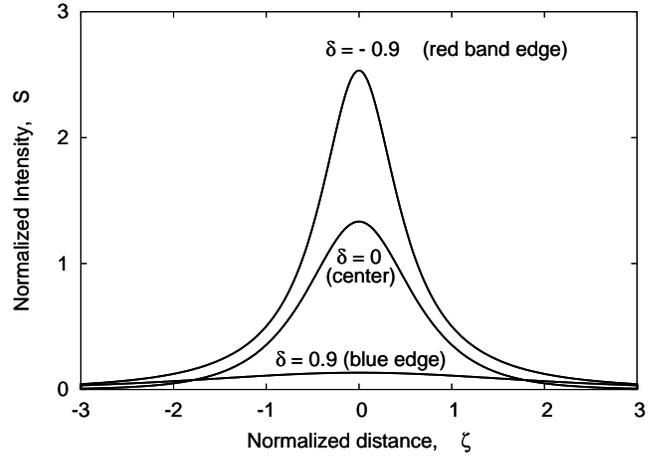}
\caption{Stationary solitons at various 
detunings $\delta$ ($\delta^2 < 1$)}
\label{fig1}
\end{figure}
(see Fig. 1, lower curve):
\begin{equation}
S  \approx { {4 ( \Delta \delta ) / 3}  \over   
{\cosh 2 ( \zeta \sqrt {2 \Delta \delta  } ) }} ;
\ \ \ \
\Phi  \approx \sqrt { 2 \Delta \delta }
\tanh ( \zeta \sqrt { 2 \Delta \delta } )
\tag{4.3}
\end{equation}
consistent with [5],
which is a familiar soliton 
of a cubic Schr\"{o}dinger equation,
whereas at the exact Bragg resonance, 
$\delta = 0$, we have a somewhat different profile,
\begin{equation}
S = 
{ 4 / 3  \over   \cosh ( 2 \zeta )  } ;
\ \ \
\tan \Phi = \sinh ( 2 \zeta )
\tag{4.4}
\end{equation}
It is worth noting that this is a solution of
a generalized Schr\"{o}dinger equation with 
the higher order nonlinearity of the 5-th
order, $\chi^{(5)}$, whereby $n = n_0 + n_4  E^4$ [16].
Finally, at the red edge, $0 < 1 + \delta \ll 1$,
the main body of a standing soliton (with the maximum peak intensity)
has the most unusual, Lorentzian, profile:
\begin{equation}
S \approx 
( 8 / 3 ) / ( 1 +   4 \zeta^2 ) ;
\ \ \
\tan \Phi \approx 4 \zeta / ( 1 + 4 \zeta^2 )
\tag{4.5}
\end{equation}
similar to a limiting case
of stimulated Raman scattering solitons [12].
Of course, far away from the peak, 
$\zeta^2  \gg ( 1 - \delta^2 )^{-1}$,
its profile decays exponentially,
$S  \propto  $ $\exp ( - 2  \zeta  \sqrt {1 - \delta^2}  )$.

The total power carried by a stationary soliton
in general case, i. e. for arbitrary $\delta$, is
\begin{equation}
W ( \delta ) = \int_{{-}  \infty}^{\infty} S  d \zeta =
{8 \over 3 }\tan^{-1} \sqrt {{1 - \delta}   \over {1 + \delta }}
\tag{4.6}
\end{equation}
and the soliton width in the $\zeta$-axis at the half-peak intensity is
\begin{equation}
Z_{1/2} ( \delta ) = { { \cosh^{-1} ( 2 + \delta )}   \over 
{\sqrt { 1 - \delta^2 } } } 
\tag{4.7}
\end{equation}
At $\delta > 0$, 
$\Delta \delta \equiv 1 - \delta \ll 1$ 
(the blue edge in case of Kerr-nonlinearity), 
solitons are the weakest and longest:
\begin{equation}
S_{pk} \approx {{4 ( \Delta \delta )}  \over 3} ;     \  \  \
W \approx {{4 \sqrt {\Delta \delta }}  \over  3} ;    
\  \  \ Z_{1/2} \approx 
{{\cosh^{-1} (1)} \over {\sqrt {2  \Delta \delta }}}
\tag{4.8}
\end{equation}
In the middle of linear band, $\delta = 0$, we have
\begin{equation}
S_{pk} = 4/3 ;    \ \ \ 
W =  2 \pi / 3 ;    \ \ \ Z_{1/2} =  \cosh^{-1} ( 2 )
\tag{4.9}
\end{equation}
The solitons at $\delta < 0$ and
$1 + \delta  \ll 1$ (red edge)
are the strongest and shortest ones,
albeit not by much:
\begin{equation}
S_{pk} \approx 8 / 3 ;     \  \  \
W \approx 4 \pi  / 3 ;    \  \  \ Z_{1/2} \approx 1
\tag{4.10}
\end{equation}

\section{CW eigenmodes of a ring Bragg reflector}
\label{sec6}

If Bragg-reflecting structure is pumped from one end,
an incident wave would decay as it propagates by transferring
its energy to a reflected wave
due to distributed retro-reflection [13].
As one traces both of them in the forward direction,
one would see the energy of both 
of them diminishing away from the incidence.
However, if one pumps the structure
from both ends, there could be conditions
when the combined back-and-forth retro-reflection may 
result in both waves 
sustaining their energy without changes.
In practical terms, such a double pumping
could be best attained in a ring (Sagnac)
optical waveguide resonator [17]
coupled to two feeding waveguides
delivering counter-propagating pumping.

Thus another objects of interest
here are what we call $cw$ "eigenmodes",
whereby both counterpropagating wave do not change as they propagate,
i. e. there is {\it no energy exchange} between them,
similarly to the eigenmodes in 
$\chi^2$ and $\chi^3$ nonlinear wave propagation [18,19].
For Bragg reflection, 
those eigenmodes exist in both linear and nonlinear
case and are allowed only {\it outside}
the Bragg bandgap (which becomes intensity-dependent 
in nonlinear medium).
A very interesting aspect of such a system 
is that Bragg eigenmodes should coexist with a 
resonant eigenmodes of a Sagnac resonator,
thus creating an interesting interplay of resonant effects.

The eigen-waves propagate with a constant phase velocity,
i. e. in eq (1.6) we can assume
$a_j  =  u_j e^{{i} q  \zeta }$
where the amplitudes $ u_j$, 
and (unknown yet) normalized perturbation
of the wave number $q$ are real constants.
[Note that there is no need to assume 
time dependence in (5.1), since the detuning
of the pumping frequency has already been taken care for by 
$\Delta$ in (1.4) or $\delta$ in (1.6).]
We will be seeking then for a 
dispersion equation between $q$ and $\delta$
(which in nonlinear case should also depend on
the wave intensities $u_j^2$)
by substituting the solution 
$a_j  =  u_j e^{{i} q  \zeta }$ into (1.6),
and derive that equation as:
\begin{equation}
q = \sqrt { ( \delta +  3 S / 2 )^2  - 1  }  -  
( u_1^2 -  u_2^2 ) / 2 , \ \ \
S = u_1^2 +  u_2^2
\tag{5.1}
\end{equation}
In linear case, $ S \ll 1$, we have
\begin{equation}
q = \sqrt { \delta^2  - 1  } ,      
\qquad\mbox{if}\quad      \delta^2  \ge 1 .
\tag{5.2}
\end{equation}
The ratio of the amplitudes of both waves is
\begin{equation}
u_j / u_{{3} - j} =  ( - 1 )^j \sqrt {
( \delta  +  3 S / 2 )^2 - 1}  - 
( \delta  +  3 S / 2 )
\tag{5.3}
\end{equation}
so that in linear case,
\begin{equation}
u_j / u_{{3} - j} =  ( - 1 )^j \sqrt { \delta^2 - 1}  -  \delta .
\tag{5.4}
\end{equation}
Both waves here have the same frequency,
$\omega  =  \omega_B ( 1  +  \mu \delta )$,
but their wavenumbers are different,
$k_j = k_B [ 1 -  ( - 1 )^j q \mu ]$.
However, in a linear case this doesn't 
amount to non-reciprocity, since
both forward and backward waves are coupled,
producing a standing-like wave
in the frame that moves toward
the higher intensity wave with "k-Doppler" velocity
${\bf v}_{Dp} =$ $\omega_B 
( k_2^{-1} -$ $k_1^{-1} ) / 2 \approx$ 
$ q \mu c / n $.
In particular, at critical points, $\delta^2 = 1$,
we have $q = 0$, hence ${\bf v}_{Dp} = 0$, 
which indicates a regular standing wave
with $u_1^2 =  u_2^2$.
In a nonlinear case, an intensity induced non-reciprocity,
the same as in a regular Kerr-line nonlinearity [13]
is due to the term $u_1^2 - u_2^2$ in (5.1).

In a nonlinear case, eq (5.3) suggests a
nonlinear connection between the eigen-waves amplitudes, $u_j $,
and the frequency detuning of a respective eigen-mode
solution. Indeed, for any given set of amplitudes $u_j $,
the respective "eigen"-detuning is:
\begin{equation}
\delta_{eigen} = - S  [ 1 + ( u_1 u_2 )^{-1} ] / 2
\tag{5.5}
\end{equation}
which, in a linear case,  $u_j^2 \ll 1$,
is consistent with (5.4), written as
$\delta_{eigen} = -  S / ( 2 u_1 u_2 ) =$ $
- ( u_1 / u_2  + u_2 / u_1 ) / 2$.
Notice that those eigenmodes in linear case
happen only outside the Bragg bandgap (5.3);
in nonlinear case they  also 
determine the boundaries of a new, 
nonlinearly modified band
whereby $ u_1^2 = u_2^2$ and 
$S = 2  u_1 u_2 $:
\begin{equation}
\delta_{cr} =  \pm 1 -  3 S / 2 ,
\tag{5.6}
\end{equation}
which is red-shifted by  $3 S / 2$ compared to
the linear one if $n_K > 0$, and blue-shifted if $n_K < 0$.

\section{Discussion}
\label{sec7}

The calculations in the previous section 
are true not only for an infinitely long fiber, 
but also for a ring fiber.
(A latter case, however, would 
involve eigen-frequencies set up by the ring length.)
A ring makes an easy experimental setup 
for the observation of those eigenmodes
by using regular external pumping fibers
coupled to the ring fiber.
Such a ring system and its eigenmodes 
(both linear and nonlinear)
may present a considerable interest
for applications related to Sagnac
effect and ring gyro based on it,
since one my expect a great enhancement
of the Sagnac effect and hence
of the sensitivity of a laser fiber-based gyro.

We didn't discussed here the stability of those
eigenmodes in strongly nonlinear regime, 
which can be readily analyzed  
using small perturbation approach,
see e. g. [20], similar to the one 
used in the theory of modulation instability.
It is worth noting however that the 
Bragg nonlinear reflection is
known to show optical bistability [21,1,2];
our preliminary research to be published
elsewhere indicates that above some critical
pumping the system is actually prone to 
{\it highly multi-stable and multi-hysteretic behavior}
by forming many quasi-solitons
inside a finite-length fiber
and switching from a $N$ stationary-soliton
mode to a $N \pm 1$ mode,
similar to the propagation of strong light in
(linearly overdense) plasma layers [10].

Furthermore, self-trapped slow or stationary solitons
in a fiber may be used for the energy storage 
in computer applications, whereby
they can be controlled to switch
from a stationary state to a high-speed
mode and used for operational memory or logic operations.
For that purpose, it may also be of interest
to use e. g. $Erbium$, $Neodimium$, 
or $Ytterium$ doped amplifying fibers or plane waveguides [17]
to get them into lasing (due to distributed retro-reflection) 
-- and at that slow-soliton supporting -- mode.

Another attractive line of further research 
would be to move from 1D-fibers 
to nonlinear 3D-photon crystals,
which most likely might be instrumental
in attaining slow or stationary 3D-solitons
$via$ the 3D-Bragg-trapping of light and 
realizing sort of "stopped light bullet"
that would greatly advance and transform 
the phenomenon of a so called "light bullet"
(we are referring here to a predicted by 
Silberberg [22] self-sustained field 
object moving with the speed of light 
of the trapping medium, and recently 
observed experimentally [23]).
In this respect, it is encouraging 
that at least a 2D-trapping
in the cross-section of light beam
in the form of random-phase gap solitons
has been observed in photonic lattices [24].

Finally, going beyond a regular Bragg-reflection
from a spatially-periodic structure, 
one can hypothesize that a similar phenomena 
of slow or stationary solitons might be expected in $any$
scattering (but low-absorbing and nonlinear)  system,
most of all -- in those with 
{\it stochastic} scattering.
Such a system may have a "washed-out"
bandgap edges, yet the main physical factors
would remain the same -- distributed
retro-reflection of light forming
quasi-standing wave and 
facilitating slow-soliton formation $via$ nonlinearity.
Those properties put it
somewhere in between Bragg-reflector 
and slightly-overdense plasma [10].

Same as above, this idea can be further
advanced by using amplifying yet scattering media,
such as doped materials.
There is a reasonable possibility 
that a strongly-scattering nonlinear
system may be able to support 
3D-self-trapping and to sustain  
of a long-lived almost unmoving 3D-"hot-ball" 
reminiscent of a ball-lighting,
provided there is a sufficient influx of 
energy from outside to support 
inverse population and 
amplification in the system.

\section{Conclusion}
\label{sec8}

A general "all-band" intensity profile of
slow-moving and stationary
bandgap (Bragg) solitons for arbitrary parameters of
of nonlinearity and spacial modulation
is found to be similar to the profiles of
$2 \pi$ solitons of cascade stimulated Raman scattering,
including "Lorentzian" solitons.
Nonlinear "no-energy exchange" eigen-modes 
may become a promising tool in exploration
and applications of such systems.


\begin{thebibliography}{99}
%
\bibitem{#1. 1D_pioneering_Bragg_Gap_solitons}
W. Chen and D. L. Mills, Phys. Rev. Lett. 58, 160 (1987);
D. L. Mills and S. E. Trullinger, Phys. Rev. B 36, 947 (1987).
%

\bibitem{#2. Schroedinger_eqn_solitons}
J. E. Sipe and H.G.Winful, Opt. Lett. 13, 132 (1988);
C.M. de Sterke and J. E. Sipe, Phys. Rev. A 38, 5149 (1988),
also in Progress in Optics, $\bf XXXIII$, 203 (1994).
%

\bibitem{#3. Hasegava_fiber-Soliton}
A. Hasegava and F. D. Tappert, Appl. Phys. Lett. $\bf 23$, 142 (1973)
%

\bibitem{#4. Zakharov}
V. E. Zakharov and A. B. Shabat, Sov. Phys. JETP $\bf 34$, 62 (1972).
%

\bibitem{#5. first_SLOW_gap_soliton_theory}
D. N. Christodoulides and R. I. Joseph, Phys. Rev. Lett.
62, 1746 (1989).
%
\bibitem{#6, Optical Solitons}
C.M. de Sterke, B. J. Eggleton, and J. E. Sipe, in
"Spatial Solitons", edited by S. Trillo and W. Torruellas (Springer-
Verlag, Berlin, 2001);
Y. S. Kivshar and G. P. Agraval, {\it Optical Solitons},
Academic Press, New York (2003), ch. V.
%

\bibitem{#7, first_experiments_slow}
B. J. Eggleton, R. E. Slusher, C. M. de Sterke, P. A. Krug, 
and J. E. Sipe, Phys.  Rev. Lett. 76, 1627 (1996);
B. J. Eggleton, C. M. de Sterke, and R. E. Slusher, 
J. Opt. Soc. Am. B 16, 587 (1999);
D. Taverner, N. G. R. Broderick, D. J. Richardson, 
R. I. Laming, and M. Ibsen, Opt. Lett. 23, 328 (1998);
P. Millar, R. M. De La Rue, T. F. Krauss, 
J. S. Aitchison, N. G. R. Broderick, and D. J. Richardson, 
Opt. Lett. 24, 685 (1999).
%
\bibitem{#8. first_immobile_Bragg}
D. Mandelik, R. Morandotti, J. S. Aitchison, and Y. Silberberg, 
Phys. Rev. Lett.  92, 093904 (2004).
%
\bibitem{#9. Bose-Einstein-gap_soliton}
O. Zobay, S. P\"{o}tting, P. Meystre, and E. M. Wright,
Phys. Rev. A, 59, 646 (1999);
B. Eiermann, Th. Anker, M. Albiez, M. Taglieber, P. Treutlein, 
K.-P. Marzlin, and M. K. Oberthaler
Phys. Rev. Lett. 92, 230401 (2004). 
%Bright Bose-Einstein Gap Solitons of Atoms with Repulsive Interaction
%
\bibitem{#10, Kaplan,plasma}
%"Optical multi-hysteresises and quasi-solitons in nonlinear plasma"
A. E. Kaplan, Optics Express, {\bf 21}, 13134 (2013).
%
\bibitem{#11, full_gap_sloton_slution}
A. B. Aceves and S.Wabnitz, Phys. Lett. A 141, 37 (1989).
%
\bibitem{cit1}
A. E. Kaplan,
%"Subfemtosecond Pulses in Mode-locked " 2 pi"-Solitons 
%of the Cascade Stimulated Raman Scattering", 
Phys. Rev. Lett. $\bf 73$, 1243 (1994);
A. E. Kaplan and P. L. Shkolnikov,
JOSA B, $\bf 13$, 347 (1996).
%
\bibitem{nonreciprocity}
A. E. Kaplan and P. Meystre, 
Opt. Lett., $\bf 6$, 590 (1981), and
Opt. Comm., $\bf 40$, 229 (1982).
%
\bibitem{14, Yariv}
H. Kogelnik and C. V. Shank, J. Appl. Phys. $\bf 43$, 2327 (1972);
A. Yariv, IEEE J. Quant. Electr., $\bf QE-9$, 919 (1973).
%
\bibitem{cit15}
A. E. Kaplan and C. T. Law, 
%``Isolas in four-wave mixing optical bistability," 
IEEE J. Quant. Electr., $\bf QE-21$, 1529 (1985).
%
\bibitem{cit16}
%``Bistable solitons"
A. E. Kaplan, Phys. Rev. Lett. $\bf 55$, 1291 (1985).
%
\bibitem{double-pumped fiber ring resonator}
J. Haavisto and G. A. Pajer, Opt. Lett. $\bf 5$, 510 (1980);
Hsien-kai Hsiao and K. A. Winick, Opt. Express, 15, 17783 (2007).
%
\bibitem{cit17}
%eigen-modes,  ``Light-induced nonreciprocity, 
% field invariants and nonlinear eigen-polarizations"
%and nonlinear eigen-polarizations," 
%"Eigenmodes of k(2) wave-mixings: cross-induced 2-nd order nonlinear refraction", 
A. E. Kaplan, Opt. Letts, $\bf 8$, 560 (1983) and $\bf 18$, 1223 (1993).
%
\bibitem{18; egenmodes for x2 and x3}
S. Trillo and S. Wabnitz, Opt. Lett. $\bf 17$, 1572 (1992).
%
\bibitem{cit17, stability of CW}
H. G. Winful, R. Zamir, and S. F. Feldman, 
%"Modulational instability in nonlinear periodic structures: implications for gap solitons," 
Appl. Phys. Lett., $\bf 58$, 1001 (1991);
A. B. Aceves, C. De Angelis, and S. Wabnitz,
Opt. Lett., $\bf 17$, 1566 (1992).
%
\bibitem{cit21, optical bistability} 
%"Theory of bistability in nonlinear distributed feedback structures," 
H. G. Winful, J. H. Marburger, and E. Garmire, 
Appl. Phys. Lett., $\bf 35$, 379 (1979).
%
\bibitem{cit1}
%(light bullet prediction) "Collapse of optical pulses,"
Y. Silberberg, Opt. Lett. $\bf 15$, 1282 (1990).
%
\bibitem{cit1}
%(light bullet observation)
S. Minardi, F. Eilenberger, Y. V. Kartashov, A. Szameit,
U. Röpke, J. Kobelke, K. Schuster, H. Bartelt, S. Nolte,
L. Torner, F. Lederer, A. T\"{u}nnermann, and T. Pertsch,
Phys. Rev. Lett. 105, 263901 (2010).
%
\bibitem{cit1}
%Observation of random-phase gap solitons in photonic lattices
G. Bartal, O. Cohen, O. Manela, M. Segev,
J. W. Fleischer, R. Pezer and H. Buljan,
Opt. Letts., $\bf 31$, 483 (2006).
%
\end{thebibliography}
\end{document}